\documentclass[journal=achemso,manuscript=article]{achemso}
\mciteErrorOnUnknownfalse
\usepackage[utf8]{inputenc}
\usepackage[T1]{fontenc}
\usepackage{lmodern}

\usepackage{amsmath, amssymb}
\usepackage{pifont}
\usepackage{graphicx}
\usepackage{siunitx}
\usepackage[flushleft]{threeparttable}
\usepackage{graphicx}
\usepackage{xcolor}
\usepackage{mathrsfs}
\usepackage{siunitx}
\DeclareSIUnit\angstrom{\text {Å}}
\DeclareUnicodeCharacter{2212}{\textminus}
\usepackage{txfonts}
\newcommand{\mrm}[1]{\ensuremath{\mathrm{#1}}}
\newcommand{\apjl}[1]{ApJL}
\newcommand{\araa}[1]{ARAA}
\newcommand{\apj}[1]{ApJ}
\newcommand{\aap}[1]{A\&A}
\newcommand{\mnras}[1]{MNRAS}
\newcommand{\jcp}[1]{JCP}
\newcommand{\apjs}[1]{ApJSS}
\newcommand{\pasp}[1]{PASP}
\newcommand{\aj}[1]{AJ}
\newcommand{\jmolspec}[1]{J. Mol. Spectrosc.}

\DeclareUnicodeCharacter{2212}{\textminus}

\usepackage{setspace}
\title{High deuteration of methanol in L1544\footnote{Based on observations carried out with the IRAM 30 m telescope. IRAM is supported by INSU/CNRS (France), MPG (Germany), and IGN (Spain).}}

\author{Silvia Spezzano$^1$, Wiebke Riedel$^1$, Paola Caselli$^1$, Olli Sipil\"a$^1$, Yuxin Lin$^1$, Hayley A. Bunn$^1$, Elena Redaelli$^2$, Laurent H. Coudert$^3$, Andr\'es Meg\'ias$^4$, Izaskun Jimenez-Serra$^4$}  

\affiliation{$^1$Max-Planck-Institut f\"ur Extraterrestrische Physik, Giessenbachstrasse 1, 85748 Garching, Germany; 
$^2$European Southern Observatory, Karl-Schwarzschild-Strasse 2, 85748 Garching, Germany; $^3$Institut des Sciences Mol\'eculaires d’Orsay (ISMO), CNRS, Universit\'e Paris-Saclay, F-91405 Orsay, France; $^4$Centro de Astrobiolog\'ia (CAB), CSIC-INTA, Carretera de Ajalvir, km 4, 28805, Torrej\'on de Ardoz, Spain}
\email{spezzano@mpe.mpg.de}

\begin{document}

\maketitle

\begin{abstract}
Isotopic fractionation is a very powerful tool to follow the evolution of material from one stage to the next in the star-formation process. Pre-stellar cores exhibit some of the highest levels of deuteration because their physical conditions (T $\leq$ 10 K and $n$(H$_2$) $\geq$ 10$^5$ cm$^{-3}$) greatly favor deuteration processes. Deuteration maps are a measure of the effectiveness of the deuteration across the core, and they are useful to study both the deuteration as well as the formation mechanism (either in the gas-phase or on grain surfaces) of the main species.
Methanol is the simplest O-bearing complex organic molecule (COM) detected in the interstellar medium (ISM). It represents the beginning of molecular complexity in star-forming regions, thus a complete understanding of its formation and deuteration is a necessary step to understand the development of further chemical complexity.
In this paper, we use single-dish observations with the IRAM 30 m telescope and state-of-the-art chemical models to investigate the deuteration of methanol towards the prototypical pre-stellar core L1544. We also compare the results of the chemical models with previous observations of deuterated methanol towards the pre-stellar cores HMM1 and L694-2.
The spectra extracted from the CHD$_2$OH map show that the emission is concentrated in the center and towards the north-west of the core. Using deep observations towards the dust and the methanol peaks of the core, we derive a very large deuterium fraction for methanol ($\sim20\%$) towards both peaks. The comparison of our observational results with chemical models has highlighted the importance of H-abstraction processes in the formation and deuteration of methanol. 
Deep observations combined with state-of-the-art chemical models are of fundamental importance in understanding the development of molecular complexity in the ISM. Our analysis also shows the importance of non-LTE effects when measuring the D/H ratios in methanol. 

\end{abstract}

\section{1. Introduction}
\label{intro}
Low-mass stars are formed by the collapse of dense cores within filamentary structures in molecular clouds \cite{bergin07, hacar23}. Dense cores are therefore crucial players in our understanding of the physical and chemical conditions at the dawn of star-formation. Pre-stellar cores are dynamically evolved starless cores with centrally concentrated density profiles and central densities higher than a few 10$^5$ cm$^{-3}$ \cite{crapsi05, keto08}. Pre-stellar cores are of particular importance in the quest of understanding the initial conditions of low-mass star-formation because they are unstable against gravitational collapse, and hence will certainly form a protostellar system. Conversely, starless cores that are less dense and not centrally concentrated (e.g. B68 and TMC-1 \cite{bergin02,fuente19}), might eventually evolve into a pre-stellar core and finally form a protostar, or dissolve back into the interstellar medium. 

The density structure of pre-stellar cores is generally modelled with a Bonnor-Ebert (BE) sphere \cite{bonnor58,ebert55} with a central plateau and a density decrease outwards that scales with $r$$^{-2}$ where the size of the central plateau decreases as the core approaches the protostar formation \cite{keto10}. Pre-stellar cores are also characterized by a steep decrease of temperatures towards their center, where the gas reaches temperatures of 6-8\,K \cite{crapsi07, pagani07}.
As a consequence of the low temperatures and high densities in the centre, molecules readily freeze onto dust grains \cite{caselli99, bacmann02, caselli22}, a process that significantly enhances deuterium fractionation \cite{crapsi05, caselli02}.
H$_2$D$^+$ and the other deuterated isotopologues of H$_3^+$ are the primary sources of deuteration in pre-stellar cores. They form $via$ the exothermic reaction

\begin{equation} 
\mathrm{H_3^+} + \mathrm{HD}  \rightarrow \mathrm{H_2D^+} + \mathrm{H_2} \\%+230~\mathrm{K} \\
\end{equation}

\noindent
 that strongly favors H$_2$D$^+$ production at temperatures below 30 K. Furthermore, the abundance of H$_2$D$^+$ is also influenced by the $ortho$-to-$para$ ratio of H$_2$, as the reverse reaction becomes endothermic when H$_2$ is predominantly in the $para$ form \cite{pagani92}. 
H$_2$D$^+$ is further deuterated by successive reactions with HD, leading to an enhancement of D$_2$H$^+$ and D$_3^+$ \cite{caselli19}. The deuterated isotopologues of H$_3^+$ are the key players in the gas-phase deuteration, while deuterium atoms, formed from the reactive dissociation of deuterated H$_3^+$ isotopologues with electrons, drive the deuteration on the icy surface of dust grains. Overall, very high levels of deuteration have been observed in pre-stellar cores, where even multiply deuterated molecules are routinely observed, e.g. $c$-C$_3$D$_2$, D$_2$CO, and CHD$_2$OH \cite{spezzano13, bacmann03, lin23a}.

Complex organic molecules (COMs) are defined as organic molecules with more than five atoms (e.g. Herbst $\&$ van Dishoeck 2009\cite{herbst09}). COMs have been observed in a wide variety of astrophysical environments. In low-mass star-forming regions, they are particularly abundant around protostars, in regions called hot corinos where forming stars heat the surrounding material above the sublimation temperature (100 K) of the water ice mantles on dust grains \cite{ceccarelli07}. In the past decade, many observations of COMs toward starless and pre-stellar cores demonstrated that they efficiently form also in very cold environments \cite{oberg10, bacmann12, vastel14, izaskun16, izaskun21, scibelli21}.
According to astrochemical models, COMs in pre-stellar core centres are mainly present in solid-phase within the thick icy mantles of dust grains (e.g. Vasyunin et al. 2017\cite{anton17}), while observable levels of COMs are found in the outskirts of pre-stellar cores (e.g. Jim\'enez-Serra et al. 2016 and 2021\cite{izaskun16, izaskun21}).
Lin et al. (2023)\cite{lin23a} recently reported on the first detection of doubly deuterated methanol (the simplest COM) towards pre-stellar cores, and derived a D/H ratio consistent with measurements in more evolved Class 0/I objects and comet 67P/Churyumov-Gerasimenko\cite{dro21}, suggesting a chemical inheritance from the pre-stellar stage. There is observational evidence suggesting that the chemical budget present in the pre-stellar phase doesn't undergo a full reset during protostar formation \cite{mueller22}. Consequently, pre-stellar cores act as a chemical reservoirs, supplying crucial building blocks for stars and planets. To follow the evolution of pre-stellar material from one evolutionary stage to the next in the star-formation process, isotopic fractionation proves to be an exceptionally powerful diagnostic tool\cite{ceccarelli14}.
 It is, in fact, not possible to reproduce the deuterium fractionation observed in water within the Solar System without taking into account the formation and deuteration of water in the pre-stellar phase \cite{cleeves14}. Furthermore, recent observations suggest that a fraction of the complex organic molecules (COMs) observed towards protostellar cores are inherited from the pre-stellar phase. \cite{vangelder20, scibelli21}. %Recent observations of singly and doubly deuterated methanol in a cometary coma of the comet 67P/Churyumov-Gerasimenko using Rosetta-ROSINA data suggest that cometary methanol stems from the pre-stellar core that birthed our Solar System \cite{dro21}.\\
Deuteration maps are a measure of the effectiveness of the deuteration across the core, and they are useful to study both the deuteration as well as the formation of the main species \cite{redaelli19, chacon19, giers22}. Furthermore, deuteration maps from multiply deuterated isotopologues (i.e. CHD$_2$OH or $c$-C$_3$D$_2$) are crucial to assess the effects of the spatial distribution of both deuterated and non-deuterated isotopologues on the deuteration peak that results from using the main isotopologue (e.g. CH$_2$DOH/CH$_3$OH).
A clear example is the deuteration of $c$-C$_3$H$_2$ observed in the pre-stellar core L1544. While the $c$-C$_3$HD/$c$-C$_3$H$_2$ column density ratio peaks at the east of the dust emission peak, the $c$-C$_3$D$_2$/$c$-C$_3$HD peaks towards the dust peak (see Figure 2 in Giers et al. 2022\cite{giers22}). This might suggest that the $c$-C$_3$HD/$c$-C$_3$H$_2$ peak could be a consequence of the steep decrease of the $c$-C$_3$H$_2$ towards the North-East in the outer layers of the pre-stellar cores, rather then a location of enhanced deuteration. In Spezzano et al. (2016)\cite{spezzano16} we showed that the southern part of the pre-stellar core L1544 is more exposed to the interstellar radiation field (ISRF), and therefore this is where the carbon chain molecules peak. The north-eastern part of the core is more shielded, and as a consequence more carbon will be locked in CO and is not available for the formation of carbon chain molecules. Given that methanol is directly formed from CO on grains \cite{watanabe02}, methanol peaks in the North-East of L1544. The deuterated isotopologues of $c$-C$_3$H$_2$, instead, are only present in the inner layers of L1544, and their distribution is not affected by the ISRF.\\
Observing multiply deuterated molecules is very important to fine-tune our astrochemical models and allow quantitative comparison among the different evolutionary stages in the star- and planet-formation process. 
With methanol being the simplest O-bearing COM and the starting point of molecular complexity in star-forming regions \cite{garrod06, chuang17}, understanding its deuteration in pre-stellar cores will provide crucial constraints on its formation and inheritance in the star-formation process. 
Although the formation of methanol on dust grains is well-established \cite{watanabe02, anton17}, the chemical pathways responsible for its deuteration remain unclear, with potential pathways including H-D substitution and hydrogenation of deuterated formaldehyde \cite{hidaka09}.
In an effort to identify crucial chemical and physical parameters for the formation and deuteration of methanol in the pre-stellar phase, Riedel et al. (2023) and (2025)\cite{riedel23, riedel25} updated a gas-grain chemical code by including various processes such as reactive desorption, diffusion mechanisms for hydrogen and deuterium atoms on the surface of interstellar dust grains, and nondiffusive reaction mechanisms. Such processes are very important to reproduce the observations of COMs in pre-stellar cores. \\
In this paper we explore the deuteration of methanol towards the prototypical pre-stellar core L1544. 
This core, located in the Taurus molecular cloud at 170 pc \cite{galli19}, is one of the best studied pre-stellar cores. Its central density is $\sim$10$^6$ cm$^{-3}$ and the central temperature is $\sim$ 6 K \cite{crapsi07}. The core exhibits a high degree of CO freeze-out and a high level of deuteration towards its center \cite{caselli99,crapsi05}. It is chemically rich\cite{vastel14, izaskun16}, showing spatial inhomogeneities in the distribution of molecular emission \cite{spezzano17}. For decades, L1544 has been the test bed for studies that have significantly advanced our understanding of the dynamical evolution of dense cores prior to star-formation. \\
The paper is structured as follows: Section 2 presents the observations, the analysis of the single-dish observations is presented in Section 3. We use state-of-the art chemical models to reproduce the deuteration of methanol in three pre-stellar cores and our results are described in Section 4. We discuss the overall results in Section 5 and summarize our conclusions in Section 6.

\begin{figure*}[!ht]

\begin{center}
\includegraphics[width=16.5cm]{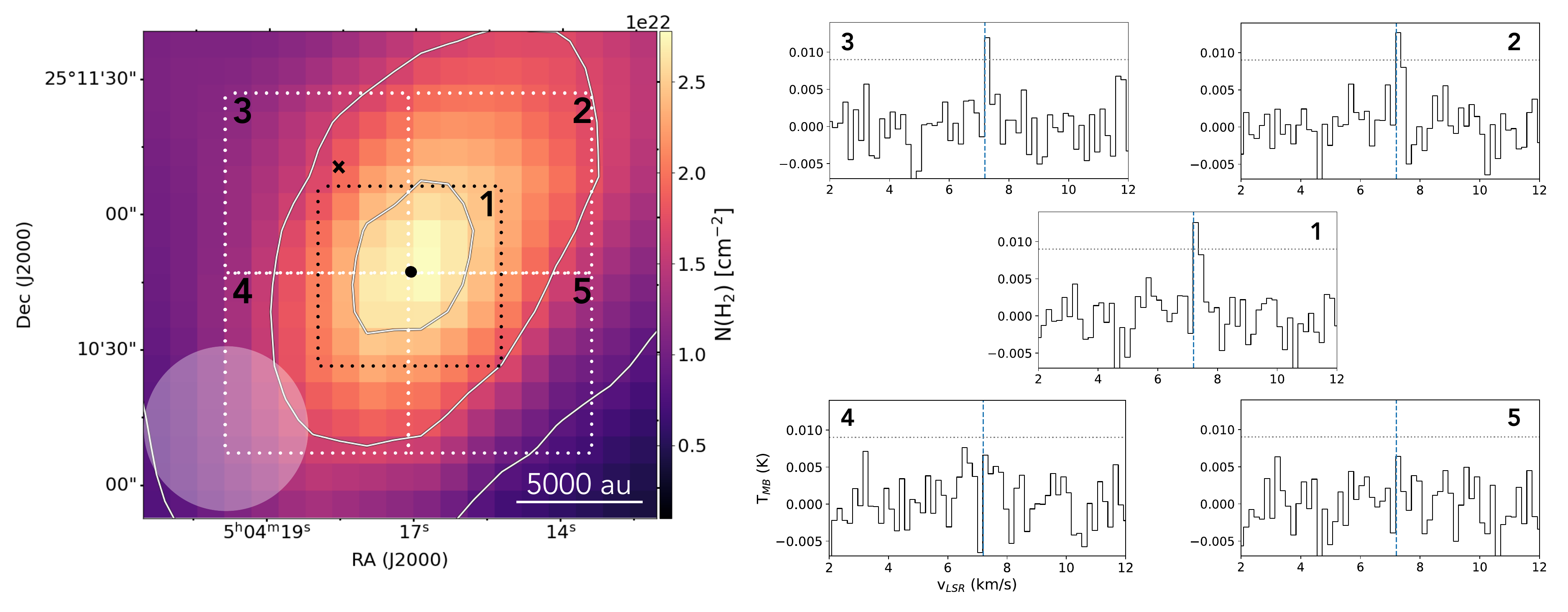}% This is a *.eps file
\end{center}
\caption{Left panel: the five different areas where the CHD$_2$OH spectra have been extracted from the OTF map observed with the IRAM 30 m telescope are shown as dotted squares on the H$_2$ column density map of L1544  computed from $Herschel$/SPIRE data at 250, 350 and 500 $\mu$m\cite{spezzano16}. The solid white contours are the 30$\%$, 60$\%$ and 90$\%$ of the peak intensity of the N(H$_2$) map. The $Herschel$/SPIRE beam is shown in the bottom left of the map. The black saltire shows the position of the methanol peak and the full black circle shows the dust emission peak. Right panel: $J$$_{K_a,K_c}$ = 2$_{0,2}$-1$_{0,1}$ $e_{0}$  CHD$_2$OH spectra extracted from the IRAM 30m OTF map. The vertical dashed lines shows the v$_{LSR}$ of the source (7.2 km/s), and the horizontal dotted lines show the 3$\sigma$ noise level. The number in each spectra refers to the area where the spectra was extracted from, shown in the left panel of the figure.}
\label{fig:map}
\end{figure*}

\section{2. Observations}
\label{observations}
\begin{figure}[!ht]
\begin{center}
\includegraphics[width=9cm]{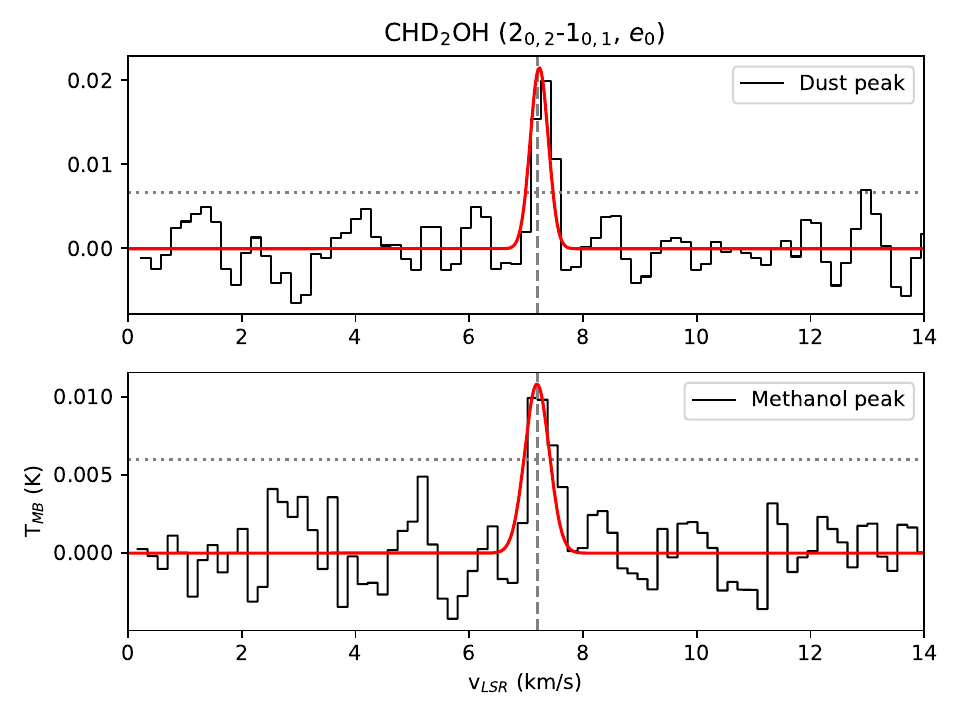}% This is a *.eps file
\end{center}
\caption{Spectra of CHD$_2$OH observed with single-pointing observations toward the dust peak (upper panel) and the methanol peak (lower panel) of L1544. The vertical dashed line shows the v$_\mrm{{LSR}}$ of the core, 7.2 km/s. The horizontal dotted lines show the 3$\sigma$ noise level.} 
\label{fig:spectra}
\end{figure}

The emission map of the  $J$$_{K_a,K_c}$ = 2$_{0,2}$-1$_{0,1}$ $e_{0}$ transition of CHD$_2$OH ($E_{\mathrm{up}}$ = 6 K) at 83289.63 MHz \cite{coudert21} towards L1544 was obtained using the IRAM 30 m telescope (Pico Veleta, Spain) in different observing runs between 2022 and 2023 (project codes: 116-21, 043-22, 104-22, PI: S. Spezzano). 
 We performed a 1.4$^\prime$ $\times$ 1.4$^\prime$ on-the-fly (OTF) map centred on the source dust emission peak ($\alpha _{2000}$ = 05$^\mrm{h}$04$^\mrm{m}$17$^\mrm{s}$.21,  $\delta _{2000}$ = +25$^\circ$10$'$42$''$.8). We used position switching with the reference position set at (-180$^{\prime \prime}$, 180$^{\prime\prime}$) offset with respect to the map centre. The EMIR E090 receiver was used with the Fourier transform spectrometer backend (FTS) with a spectral resolution of 50 kHz. The mapping was carried out in good weather conditions ($\tau_{225~\mrm{GHz}}$ $\sim$ 0.3) and a typical system temperature of T$_{sys}$ $\sim$ 90-150 K. The data processing was done using the GILDAS software \cite{pety05}. The emission map has a beam size of 30$^{\prime\prime}$, and was gridded to a pixel size of 6$''$ with the CLASS software in the GILDAS package, which corresponds to $\sim$1/5 of the beam size. The intensity scale was converted into main beam temperature T$_{MB}$ assuming the forward efficiency F$_{eff}$ = 0.95 and B$_{eff}$ = 0.81. The noise level was homogeneous in our map, therefore no weighting was applied to the individual spectra before averaging.
 While the brightest CH$_2$DOH transition in the 3 mm band was observed, the line is still too weak to produce an integrated intensity map. The averaged spectra towards five different regions across L1544 are shown in Figure~\ref{fig:map}.
 The single pointing observations towards the dust peak of L1544 shown in Figure~\ref{fig:spectra} are from the IRAM 30 m large program ASAI \cite{lefloch18}. The single pointing observations towards the methanol peak ($\alpha _{2000}$ = 05$^\mrm{h}$04$^\mrm{m}$18$^\mrm{s}$,  $\delta _{2000}$ = +25$^\circ$11$'$10$''$) shown in Figure~\ref{fig:spectra} were obtained with the 30m telescope in 2024 within the framework of project 022-24 (PI: A. Meg\'ias).

\begin{table*}[!ht]
\centering
\caption{Parameters of the observed CHD$_2$OH lines in the dust and methanol peak of L1544.}
\label{tab:obs}
 \scalebox{0.80}{
 \begin{tabular}{cccccccc}
\hline \hline
& $W$ & $v_\mathrm{LSR}$ & FWHM & rms & $N_\mathrm{TOT}$ (T$_\mathrm{ex}$ = 5 K) & $N_\mathrm{TOT}$ (T$_\mathrm{ex}$ = 6.5 K) & $N_\mathrm{TOT}$ (T$_\mathrm{ex}$ = 8 K)\\
 & mK km s$^{-1}$ & km s$^{-1}$ & km s$^{-1}$ & mK & cm$^{-2}$ & cm$^{-2}$ & cm$^{-2}$ \\
\hline
dust peak &8.4(9)&7.19(2)&0.37(4)&3 &6.7(7)$\times10^{11}$&7.0(8)$\times10^{11}$ &8.0(9)$\times10^{11}$ \\
methanol peak &6.2(9)&7.18(5)&0.47(9)&3& 5.0(7)$\times10^{11}$ & 5.2(8)$\times10^{11}$&5.9(9)$\times10^{11}$  \\

 \hline
\end{tabular}
}
\begin{tablenotes}
\item \textit{Note:} The laboratory spectroscopy reference for CHD$_{2}$OH is Drozdovskaya et al. (2022)\cite{dro22}. The integrated intensities are reported in units of T$_{MB}$. Numbers in parentheses denote 1$\sigma$ uncertainties in units of the last quoted digit.
\end{tablenotes}
%\tablefoot{The laboratory spectroscopy reference for CHD$_{2}$OH is \citet{Dro22}. }
\end{table*}

\section{3. Results}
\label{Results}
The results of the IRAM 30m project aimed at mapping the 2$_{0,2}$-1$_{0,1}$, $e_0$ transition of doubly deuterated methanol toward the central 1.4$^\prime$ $\times$ 1.4$^\prime$ region of L1544 are shown in Figure~\ref{fig:map}. The final rms of the map is $\sim$5 mK and the peak intensity of the CHD$_2$OH line, towards the dust peak, is $\sim$
15 mK. Given the weakness of the line, we used the OTF data to average the spectra within an area of 40$^{\prime\prime}$$\times$ 40$^{\prime\prime}$ towards five quadrants shown with dotted white lines in Figure~\ref{fig:map}. All spectra in Figure~\ref{fig:map} are plotted between -0.007 and  0.012 K, and between 2 and 12 km/s, with the emission lines centered at 7.2 km/s. The color map used as background for the spectra is the N(H$_2$) map of L1544 computed from $Herschel$/SPIRE data \cite{spezzano16}. The spectra in Figure~\ref{fig:map} show that the CHD$_2$OH line is brightest in the central and the northwest part of the core, where it is detected with S/N>4 over two channels. The line is barely detected in the North-East (S/N$\sim$4) over only one channel, and not detected in the southern part of the core.
The result is in agreement with the singly deuterated methanol maps shown in the central and right panels of Figure 2 in Chac\'on-Tanarro et al. (2019)\cite{chacon19}.  \\
Although the results shown in Figure~\ref{fig:map} allow us to understand the distribution of doubly deuterated methanol in L1544 and qualitatively compare with OTF maps of CH$_3$OH, CH$_2$DOH, and other deuterated isotopologues observed in L1544 \cite{chacon19, redaelli19, giers22}, the poor signal-to-noise ratio of the spectra would make a quantitative comparison rather inconclusive. 
To overcome this limitation, we use deep observations towards the dust and methanol peaks of L1544, where the rms is 2.2 and 2.0 mK, respectively. The single pointing observations are shown in Figure~\ref{fig:spectra}. The results of the Gaussian fit towards the dust and methanol peak of L1544 are reported in Table~\ref{tab:obs}.
The column densities of CHD$_2$OH reported in Table~\ref{tab:obs} have been computed from the spectra shown in Figure~\ref{fig:spectra} using the formula reported in Mangum $\&$ Shirley (2015)\cite{Mangum15}, assuming optically thin emission and that the source fills the beam:

\begin{equation} 
 N_\mrm{{tot}} = \frac{8\pi\nu^3Q_{rot}(T_{ex})W}{c^3A_{ul}g_u}\frac{e^{\frac{E_u}{kT}}}{J(T_{ex}) -  J(T_{bg})},\\
\end{equation}

\noindent
where $J(T) = {\frac{h\nu}{k}}(\mrm{e}^{\frac{h\nu}{kT}}-1)^{-1}$ is the Rayleigh-Jeans
equivalent temperature in Kelvin, $k$ is the Boltzmann constant, $\nu$ is the frequency of the line, $h$ is the Planck constant, $c$ is the speed
of light, $A_{ul}$  is the Einstein
coefficient of the transition, $W$ is the integrated intensity, $g_u$ is the degeneracy of the upper state, $E_u$ is the
energy of the upper state, $Q_{rot}$ is the partition function of the molecule at the given temperature $T_{ex}$, $T_{bg}$is the
background (2.7 K). 
We calculated the partition function $Q(T\mathrm{_{ex}})$ at 5, 6.5, and 8 K using the CHD$_{2}$OH catalog from the CDMS \cite{mueller05}, recently updated based on Drozdovskaya et al. (2022)\cite{dro22}. The resulting partition functions are reported in Table~S1.
The column densities of doubly deuterated methanol reported in Table~\ref{table:CD} were calculated considering variations of $T_{\mathrm{ex}}$ within 5-8 K and assuming a calibration error of 20$\%$ to derive the uncertainties, as done in Lin et al. (2023)\cite{lin23a}. Table~\ref{table:CD} also reports on the deuteration ratios of methanol at the dust and methanol peaks of L1544, as well as the column densities of the main and singly deuterated isotopologues of methanol reported in Lin et al. (2022)\cite{lin22} and Chac\'on-Tanarro et al. (2019)\cite{chacon19}, for completeness. The column densities of the main isotopolog reported in Table 5 of Lin et al. (2022)\cite{lin22} have been derived with RADEX non-LTE modelling using a total of ten different lines (of which four were obeserved as upper limits) at 3 and 2 mm, and they agree within a factor of two with previous values reported in Vastel et al. 2014\cite{vastel14}, Bizzocchi et al. 2014\cite{bizzocchi14}, Punanova et al. (2018)\cite{punanova18}, and Chac\'on-Tanarro et al. (2019)\cite{chacon19}.

\begin{table*}[!ht]
\caption{Column densities and column density ratios at the dust peak and methanol peak of L1544}
\label{table:CD}
%\scalebox{1}{
\begin{tabular}{ccc}
\hline\hline \\[-2ex]
& Dust Peak & Methanol Peak   \\
\hline
$N$(CH$_3$OH)$^a$&  1.30(5)$\times$10$^{13}$ cm$^{-2}$     &     1.60(3)$\times$10$^{13}$ cm$^{-2}$        \\
$N$(CH$_2$DOH)$^b$&  2.8(7)$\times$10$^{12}$ cm$^{-2}$     &     3.3(8)$\times$10$^{12}$ cm$^{-2} $       \\
$N$(CHD$_2$OH)$^c$&    7.2(1.4)$\times$10$^{11}$ cm$^{-2}$  &     5.4(1.5)$\times$10$^{11}$ cm$^{-2}$          \\
\hline
$N$(CH$_2$DOH)/$N$(CH$_3$OH) &22(6)$\%$&21(5)$\%$\\
$N$(CHD$_2$OH)/$N$(CH$_3$OH) &6(1)$\%$&3(1)$\%$\\
$N$(CHD$_2$OH)/$N$(CH$_2$DOH) &26(8)$\%$&16(6)$\%$\\
\hline
\end{tabular}
\begin{tablenotes}
\item \textit{Note:} $^a$From Lin et al. 2022\cite{lin22}. $^b$From Chac\'on-Tanarro et al. 2019\cite{chacon19}. $^c$This work. Numbers in parentheses denote 1$\sigma$ uncertainties in units of the last quoted digit.
\end{tablenotes}
%\tablefoot{$^a$From Lin et al. 2022. $^b$From Chacon-Tanarro et al. 2019. $^c$This work. \\}
\end{table*}

\section{4. Chemical models}
\label{Models}

To compare if our current theoretical understanding of deuterium chemistry can match the observed deuteration trends for methanol, we have tested several models originally developed to reproduce the CH$_2$DOH/CH$_3$OH ratio\cite{riedel23, riedel25}.

The chemical evolution of molecular abundances is computed with the gas-grain astrochemical code \textit{pyRate} \cite{sipila15, sipila19}. The chemical network for the gas-phase is based on the 2014 public release of the Kinetic Database for Astrochemistry \cite{kida}. A recent update to the latest data release (kida.uva.2024, Wakelam et al. 2024\cite{kida24}) was tested on a 0D model using a non-deuterated chemical network, and showed no significant deviations for methanol, and hence for simplicity we decided to proceed with the existing deuterated networks\cite{riedel23, riedel25}.
The grain surface network is based on the one presented in Semenov et al. (2010)\cite{semenov10}. Reactions were cloned to include deuterated counterparts for species up to seven atoms and spin-state counterparts for selected species. Uncertainties arise when reactions are cloned to describe the evolution of deuterated species. The methanol formation and deuteration scheme follow the experimentally verified proposal by Hidaka et al. (2009)\cite{hidaka09}. Generally, experimental data is used when it is available; for details on the network and deuteration schemes, we refer to Riedel et al. (2025)\cite{riedel25} and references therein.
%The methanol formation and deuteration scheme follows the experimentally verified proposal by Hidaka et al. (2009)\cite{hidaka09}.
The models assume a three-phase grain model, including a gas-phase, a chemically active surface-phase, and an inert mantle-phase. Desorption of methanol from the surface of the dust grain occurs predominantly through non-thermal desorption mechanisms in the extremely cold conditions of pre-stellar cores. Usually, reactive desorption is presumed to be the dominant one\cite{anton17, riedel23}. All models presented in this work apply a constant reactive desorption efficiency of 1\% \cite{garrod07}. The formation enthalpies and binding energies used in the model are reported in Table A.1 of Riedel et al. (2023)\cite{riedel23}. We note that a few binding energies listed in Table A.1 of Riedel et al. (2023)\cite{riedel23} have been recently revised in theoretical and experimental studies (e.g. Minissale et al. 2022\cite{minissale22}). However, in cold environments such as L1544, binding energies exceeding $\sim$2000 K are too high for thermal or CR-induced desorption to have a substantial impact on gas-phase abundances.
After a recent update \cite{riedel25}, pyRate includes several nondiffusive reaction mechanisms. However, their impact on the chemistry of methanol, which is mainly formed and deuterated by addition and abstraction reactions of highly mobile H and D atoms, was found to be only minor in Riedel et al. (2025)\cite{riedel25}, where a factor of non-diffusive/diffusive of 1.07 (dust peak) and 0.95 (methanol peak) is derived at the best fit-time (t=3$\times$10$^5$ yr).
The models presented in this work, therefore, include solely the more conservative diffusive chemistry. Nonetheless, we note that Jimenez-Serra et al. (2025)\cite{izaskun25} found that the formation of CO, CO$_2$ and CH$_3$OH are tightly linked, so that non-diffusive chemistry may lead to some different results. Surface reactions proceed through the Langmuir-Hinshelwood mechanism relying on thermal diffusion.
Riedel et al. (2025) \cite{riedel25} tested over 30 different models while investigating the formation and deuteration of methanol in cold dense cores, like L1544. Here we compare the result of the best four models (D2, D3, D4 and D5) against our observations in L1544. To facilitate the comparison, we kept the same nomenclature as in Riedel et al. (2025)\cite{riedel25}. The characteristics of the models used in this paper are listed in Table~S2. Models D2, D3 and D4 adopt only H-addition reactions, while model D5 also includes H-abstraction reactions. %Models D2 and D3 apply the single collision model proposed by Hagesawa et al. (1992) with either tunnel diffusion (D2) or fast diffusion (D3). Models D4 and D5 apply the reaction-diffusion competition model proposed by Chang et al. (2007). 
Model D2 additionally allows for the diffusion of hydrogen and deuterium atoms by quantum tunneling through a rectangular barrier of \SI{1}{\angstrom} width. The diffusion-to-binding energy $E_{\mathrm{d}}/E_{\mathrm{b}}$ is set to 0.55; with the exception of model D3, where it is set to 0.2, the lowest value debated in the literature \cite{furuya22}.
Reactions with an activation-energy barrier play an important role in the hydrogenation (and deuteration) of methanol. Hence, the approach used to derive their reaction probabilities has a significant impact on the formation of methanol and its deuterated isotopologues. Here, we test two approaches widely used in the literature. Models D2 and D3 apply the single collision approach \cite{hasegawa92}, which assumes that the reactant has only one attempt to either thermally hop over the barrier or tunnel through it. Models D4 and D5 apply the reaction-diffusion competition approach \cite{chang07}, which considers that the reaction partners are confined in the same binding site until one of them diffuses away again and can therefore undergo multiple attempts to react with each other. The chemical models were run using the physical structure of L1544 \cite{keto15}, shown in Figure S2. All models use the initial chemical abundances reported in Table 1 of Riedel et al. (2025)\cite{riedel25}, consider a spherical dust grain with a radius of 0.1 $\mu$m and a surface density of binding sites of 1.5$\times$10$^{15}$ cm$^{-2}$. This work uses the canonical value for $\zeta$$_2$ (1.3$\times$10$^{-17}$ s$^{-1}$). We note that a recent re-evaluation has been presented in Redaelli et al. (2025)\cite{redaelli25}, and the revised value is consistent within the uncertainty of the method with the canonical value of 1.3$\times$10$^{-17}$ s$^{-1}$. The visual extinction in the models is calculated as 
$A_V = 10^{-21} \, N(\mathrm{H}_2)$; a floor value of 1 mag for L1544 and L694-2, and 3 mag for HMM1, are added to account for the more extended envelope. The external values used for the three cores are 
1~mag for L1544 and L694-2, and 3~mag for HMM1. The resulting  abundances were converted to column densities including beam convolution with a beam size corresponding to the observations. The results of the models for L1544 and the comparison with the observations are shown in Figure 3. \\

\begin{figure*}
\begin{center}
\includegraphics[width=17cm]{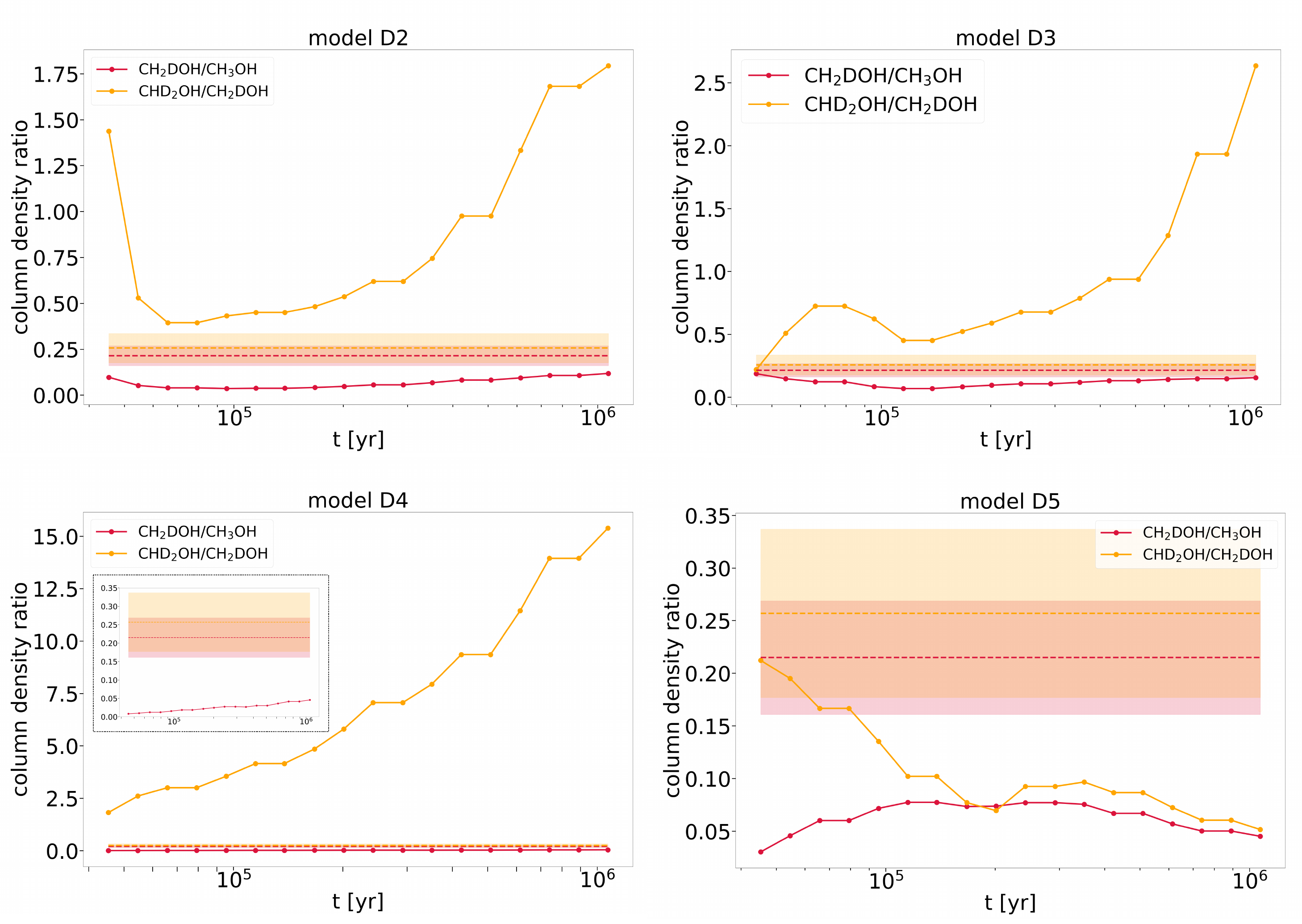}% This is a *.eps file
\label{fig:L1544_models}
\end{center}
\caption{Column density ratios for the deuteration of methanol in L1544 computed with four of the models presented in Riedel et al. (2025)\cite{riedel25}. The horizontal dashed lines show the result from the observations towards the dust peak of L1544 and the shaded region indicates the error bars of the observed ratios. Models D2 and D3 apply the single collision model proposed by Hasegawa et al. (1992)\cite{hasegawa92} with either tunnel diffusion (D2) or fast diffusion (D3). Models D4 and D5 apply the reaction-diffusion competition model proposed by Chang et al. (2007)\cite{chang07}. Additionally, D5 allows for H abstraction reactions. For model D4, a zoom-in for low values of column density ratios has been added within the plot.}
\end{figure*}

\begin{figure*}
\begin{center}
\includegraphics[width=18cm]{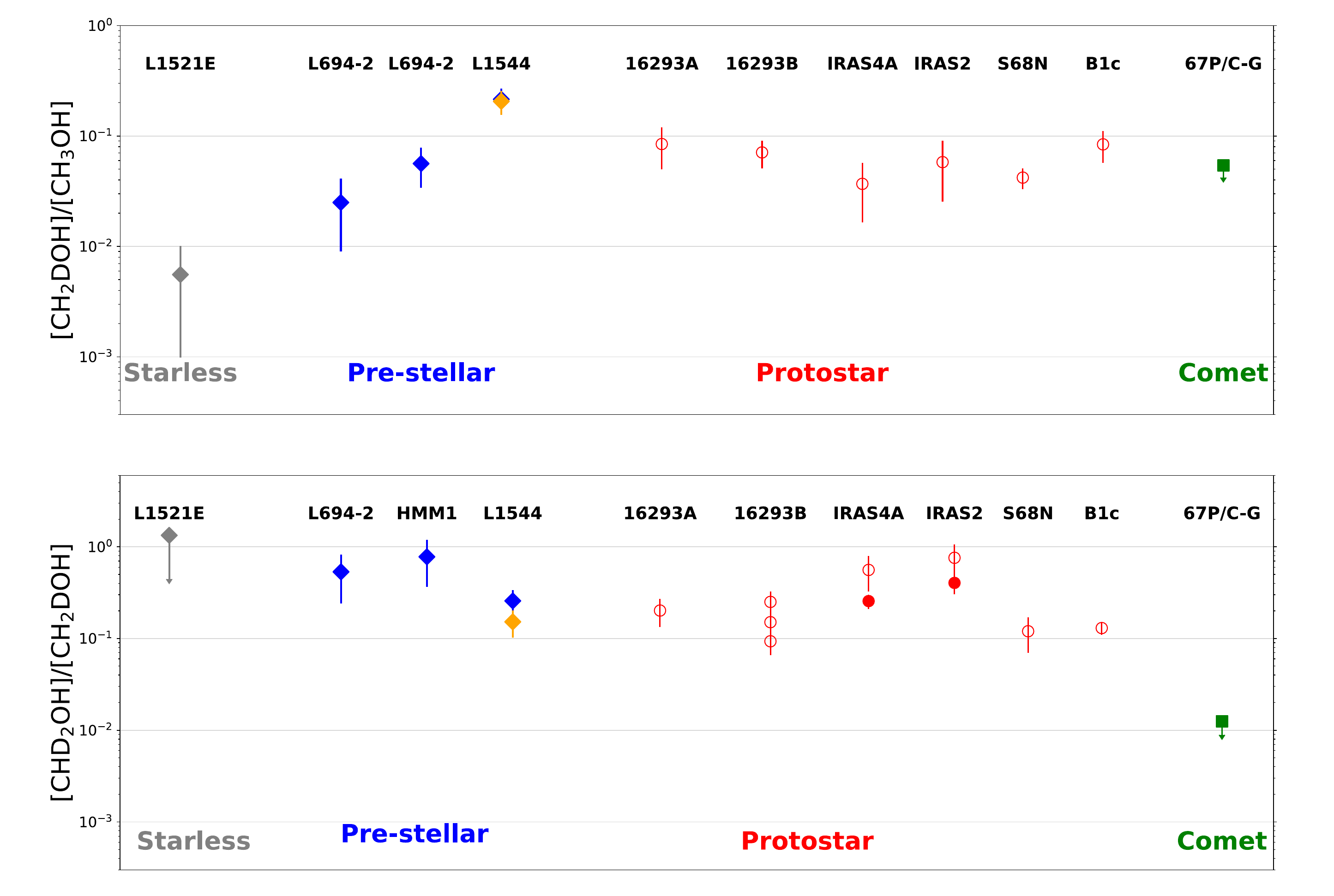}% This is a *.eps file
\end{center}
\caption{The column density ratios of [CH$_2$DOH]/[CH$_3$OH] (upper panel) and [CHD$_2$OH]/[CH$_2$DOH] (lower panel) as a function of source types, rearranged from Lin et al. (2023a)\cite{lin23a} with L1544 values in the dust (blue) and methanol (orange) peaks (from this work). Filled markers indicate single-dish observations and open markers indicate interferometric observations. The plotted values for L1544 take into consideration T$_\mrm{{ex}}$ variations from 5 to 8 K, and use the CH$_3$OH column density derived in Lin et al. (2023b)\cite{lin23b} with non-LTE models. The references for S68N and B1c are from van Gelder et al. (2022)\cite{vangelder22}; for IRAS4A and IRAS2 are Taquet et al. (2019) and Parise et al. (2006)\cite{taquet19, parise06}; for IRAS16293A and IRAS16293B are Manigand et al. (2019)\cite{manigand19}, Jørgensen et al. 2016\cite{jorgensen16}, Drozdovskaya et al. (2022)\cite{dro22}; for comet 67P/C-G is Drozdovskaya et al. (2021)\cite{dro21}.}
\end{figure*}

\section{5. Discussion}
\label{Discussion}
The spectra on the L1544 map in Figure~\ref{fig:map} show that the line of CHD$_2$OH is not detected towards the southern part of the core, and detected at a 2$\sigma$ level (in integrated intensity) towards the north-east part of the core. The distribution of methanol in L1544 is characterized by a sharp decrease towards the South because of a more efficient illumination from the interstellar radiation field \cite{spezzano16}. It is therefore not surprising that we do not observe CHD$_2$OH in the southern part of the core. 
On the contrary, it might be surprising that the line is very weak towards the north-east part of the core, given that the methanol peak is located towards the north east with respect to the dust peak of L1544 \cite{bizzocchi14}. It is important to note, however, that the quadrants that we used to average the spectra shown in Figure 1 are large and the position of the methanol peak is covered by both the central and the upper-left quadrant. Overall, the spectra in Figure 1 show that our target line for CHD$_2$OH is observed in a rather small portion of the map around and slightly towards the north of the dust peak.

When comparing the deuteration of methanol towards the dust and methanol peaks in L1544 using the deep observations shown in Figure 2, we do not see significant differences within error-bars. The deuterium fractions measured towards the dust peak, however, tend to be larger than the ones measured towards the methanol peak, as reported in Table 2. 
The deuteration maps of N$_2$H$^+$, HCO$^+$, and $c$-C$_3$H$_2$ in L1544 also peak towards the center of the core (e.g., Redaelli et al. 2019 and Giers et al. 2022\cite{redaelli19, giers22}) where the deuteration is more efficient because of the local increase in the abundance of H$_2$D$^+$, D$_2$H$^+$, and D$_3^+$, as well as the catastrophic freeze-out of CO \cite{caselli99}. The level of deuteration reached by each molecule varies, and it is likely influenced both by the molecule's distribution within the different layers of the core\cite{redaelli19, spezzano16}, as well as by the relevant deuteration processes for each molecule. N$_2$H$^+$ has larger  levels of deuteration (26\%) than $c$-C$_3$H$_2$ (17\%) and HCO$^+$ (3.5\%) because it traces best the denser gas in the centre of L1544, where the deuteration is more efficient. The level of deuteration measured in methanol is similar to N$_2$H$^+$ even if methanol traces an outer shell of L1544\cite{bizzocchi14, vastel14, spezzano16}, as $c$-C$_3$H$_2$ does. This is indicative of a much more efficient deuteration process taking place in the interstellar ices, where methanol and deuterated methanol are formed, in comparison with the deuteration happening in the gas-phase (e.g. for $c$-C$_3$H$_2$). The deuteration ratio R$_\mathrm{{D}}$=$N$(XHD)/$N$(XH$_2$) is $\sim$15\% for $c$-C$_3$H$_2$ and $\sim$20\% for methanol, while the R$_\mathrm{{D_2}}$=$N$(XD$_2$)/$N$(XHD) is $\sim$1\% for $c$-C$_3$H$_2$ and $\sim$25\% for methanol, indicating that the second deuteration of methanol, to form CHD$_2$OH, is also more efficient than the second deuteration of $c$-C$_3$H$_2$, to form $c$-C$_3$D$_2$. Particularly puzzling is the deuteration of H$_2$CO and H$_2$CS, whose large R$_\mathrm{{D_2}}$ ($\sim$100\%) towards the dust peak of L1544. While, unlike methanol, H$_2$CO and H$_2$CS can also be formed and deuterated in the gas-phase (e.g. Zahorecz et al. 2021\cite{zahorecz21}), the large  R$_\mathrm{{D_2}}$ observed in L1544 cannot be reproduced with chemical models that consider the deuteration on the surface as well as in the gas-phase \cite{chacon19, spezzano22}. 

To understand the different deuterium fractions observed in L1544, we use the best models among the ones developed and tested in Riedel et al. (2025) for L1544, as described in Section 4, and compared the results against the observed trends. The results, shown in Figure 3 clearly indicate that model D5 is the only one that does not predict very large R$_\mathrm{{D_2}}$ ratios that would strongly disagree with our observations. Additionally, model D5 predicts relatively similar values for R$_{D_2}$ and R$_{D}$, which is in agreement with our observations. This is a very interesting result because model D5 is the only one that includes the H-abstraction reactions. H-abstraction reactions have been studied in the laboratory\cite{hidaka09,chuang16,minissale16}, and the experimental results showed their importance in the reaction scheme for methanol formation.

Lin et al. (2023a)\cite{lin23a} reported on the first detection of doubly deuterated methanol towards pre-stellar cores and observed deuterium fractions towards L694-2 and HMM-1 that are different than what we observe in L1544. The R$_\mathrm{{D}}$ is 3\% in L694-2 and 6\% in HMM-1, lower than what we observe for L1544 (20\%). On the other hand, R$_\mathrm{{D_2}}$ is 50\% in L694-2 and 80\% in HMM-1, larger than what we observe in L1544 (25\%).
 Figure S1 shows the results of the chemical modelling using model D5 on L694-2 and HMM-1, with the physical structure of the core being the only difference when applying model D5 to the different cores in our sample. It is very interesting to note the effect that the different physical structures of the three cores (L1544, L694-2, and HMM1), shown in Figure S2, have on the predicted ratios. Figure S3 shows the results of the D5 models for the three cores, to facilitate the comparison among them. Additionally, it is worth noticing that the observed R$_\mathrm{{D_2}}$ and R$_\mathrm{{D}}$ ratios in L694-2 and HMM1 can also be well reproduced within a factor of 2.  \\

In Figure 4, we have included our results on L1544 to the plots shown in Figure 2 of Lin et al. (2023a)\cite{lin23a}. The summary plots in Figure 4 show the values of R$_{D}$, and R$_{D_2}$ reported in the literature for starless and pre-stellar cores, protostars, and comets. As already discussed in Lin et al. (2023a), there is a strong observational evidence that the deuteration of methanol is enhanced in dynamically evolved cores, and that the pre-stellar methanol is efficiently inherited in the protostellar phase.
R$_\mathrm{{D_2}}$ shows the least variations across the sources in Figure 4 because singly and doubly deuterated methanol are more likely to trace the same gas, while the normal isotopologue is also present in regions of the cores where deuteration is not efficient. A non-LTE analysis for the excitation of CH$_3$OH was considered for the starless and pre-stellar cores in Figure 4, while for the other objects in Figure 4 the analysis for CH$_3$OH was done under the assumption of LTE. The differences in R$_{D}$ that arise from using the LTE vs non-LTE analysis can be significant. In the case of L1544, for example, R$_{D}$ is 7(2)$\%$ assuming LTE \cite{chacon19}, while we derive here a value of 22(6)$\%$ using the column density of CH$_3$OH computed with a non-LTE analysis in Lin et al. (2022)\cite{lin22}. As a consequence, the values shown in Figure 4 for the for the protostars and the comet may differ by a factor of three. \\

\section{6. Conclusions}
\label{Conclusions}
Isotopic fractionation, and in particular deuteration, is an excellent tool to understand the formation and inheritance of molecules in star-forming regions. Towards the pre-stellar core L1544, methanol exhibits levels of deuteration as large as N$_2$H$^+$, highlighting its very efficient deuteration on the icy surface of dust grains.\\
By comparing our observational results with state-of-the-art chemical models, we are able to gauge the importance of H-abstraction reactions in the formation and deuteration of methanol on the surface of dust grains. Additionally, we have compared the observations of three pre-stellar cores and assessed the large effect that their physical structures have on the deuteration of methanol.\\
Collisional rate coefficients for deuterated methanol will be necessary to assess non-LTE effects and consequent effects on the column density that we routinely derive in star-forming regions.\\
Methanol represents the beginning of molecular complexity in star-forming regions, thus a complete understanding of its formation and deuteration is a necessary step to understand the development of further chemical complexity. In this regard, understanding the processes responsible for the very high R$_\mathrm{{D_2}}$ measured in H$_2$CO, an intermediate in the formation of methanol on the surface of dust grains, is of paramount importance.

\section*{Supporting Information}
\begin{itemize}
    \item On the spectroscopy and catalogs of deuterated methanol
    \item Dependence of deuteration on physical conditions
\end{itemize}

\section*{Acknowledgments}
The authors wish to thank the anonymous referee for the careful review of the manuscript. We gratefully acknowledge the support of the Max Planck Society. I.J-.S and A.M. acknowledge funding by the ERC CoG grant OPENS, GA No. 101125858, funded by the European Union. I.J-.S and A.M. also acknowledge support from grant PID2022-136814NB-I00 funded by the Spanish Ministry of Science, Innovation and Universities/State Agency of Research MICIU/AEI/ 10.13039/501100011033 and by “ERDF/EU”.

\bibliography{myreferences}

\newpage
{\huge Supporting Information}

\renewcommand{\thefigure}{S\arabic{figure}}
\renewcommand{\thetable}{S\arabic{table}}
\renewcommand{\thepage}{S\arabic{page}}
\setcounter{figure}{0}
\setcounter{table}{0}
\setcounter{page}{1}

%\begin{apendix}
\section{On the spectroscopy and catalogs of deuterated methanol}
The internal rotation of the asymmetric methyl group in the isotopologues of methanol with deuterium in the methyl group leads to complex spectral patterns that require complex analysis. As a consequence, particular care needs to be taken when using data from online catalogs as approximate methods used to produce the catalog might have effects on the interpretation of astronomical data.
In the JPL catalog, the rest frequencies of CH$_2$DOH transitions that are energetically favorable to observe in cold sources like starless and pre-stellar cores have errors in the order of $\sim$100 kHz with respect to the measured frequencies reported in the supplementary material of Coudert et al. (2014)\cite{coudert14}. Such small deviations are significant in cold sources because of the characteristic small line-widths and might induce significant error in the velocity of the line or even misidentification. We therefore suggest to refer to the rest frequencies reported in the spectroscopy papers and compare them to the catalogs, especially in cases of doubts on the interpretation of the astronomical data.
The complexity of the rotational ladder also translates into potential errors when extrapolating the values of their partition functions at temperatures not listed in the online catalogs. We therefore list in Table~\ref{table:Q} the partition functions Q(T) for CH$_2$DOH and CHD$_2$OH in a large range of temperatures, including temperatures relevant for cold sources like starless and pre-stellar cores. The values reported in Table~\ref{table:Q} have been computed using all torsional levels up to 1700 cm$^{-1}$ for CH$_2$DOH and 2000 cm$^{-1}$ for CHD$_2$OH. We note that the partition function reported in Table~\ref{table:Q} for CH$_2$DOH is very close to the values currently listed in the JPL catalog, despite the catalog only listing transitions from the ground torsional state based on Pearson et al. (2012)\cite{pearson12}. It is plausible that a correction factor has been used to correct the partition functions listed in the JPL catalog (Drouin, priv. comm.). This warrants a re-evaluation of the column densities derived with the JPL partition function that used a correction factor (e.g. J\o{}rgensen et al. 2018\cite{jorgensen18}).

\begin{table*}{}
\caption{Partition function, Q(T), of CH$_2$DOH and CHD$_2$OH }
\label{table:Q}
%\scalebox{1}{
\begin{tabular}{ccc}
\hline\hline \\[-2ex]
T(K) &CH$_2$DOH &CHD$_2$OH  \\
\hline
300 &16753.496 &19331.489 \\
225 &9506.242 &11256.666\\
150 &4385.771 &5255.582\\
75.0  & 1294.257&1563.391 \\
37.5  & 398.999 &490.483\\
18.75 &  114.589 &145.295 \\
9.375 &  30.388 & 39.851\\
8.0 &22.447 &29.521 \\
6.5 &15.210 &20.009 \\
5.0 & 9.603&12.412 \\
\hline
\end{tabular}
\end{table*}

\begin{table*} 
	\caption{Overview of the four models from Riedel et al. (2025)\cite{riedel25} used in this work.}
	\label{table:models}
	\begin{tabular}{l  c c c l}
		\hline
		\hline
		Model  & RD$^a$ efficiency & H$_2$ removal$^b$ & other modifications \\
		\hline
		D2$^{\star}$  & 1\% & scaled E$_{\mathrm{b}}$ & tunnel diffusion\\
		D3$^{\star}$ &  1\% & scaled E$_{\mathrm{b}}$ & fast diffusion \\
		\hline
		D4$^{\dag}$ & 1\% & scaled E$_{\mathrm{b}}$ & \\
		\hline
	D5$^{\dag}$ &  1\% & scaled E$_{\mathrm{b}}$ &  H-abstraction reac.\\
\end{tabular}
\begin{tablenotes}
\item \textit{Note:} $^{\star}$ Models that apply the single collision model proposed by Hasegawa et al. (1992)\cite{hasegawa92} to derive the reaction probabilities. $^{\dag}$ Models that apply the reaction-diffusion competition model proposed by Chang et al. (2007) \cite{chang07} to derive the reaction probabilites. $^a$ Efficiency of the reactive desorption. $^b$ To avoid unphysical build-up of H$_2$ (and its deuterated isotopologues) on the surface, their binding energies have been scaled by a factor of 0.1\cite{riedel25}.
\end{tablenotes}
\end{table*}

\begin{figure}[!ht]
\begin{center}
\includegraphics[width=9cm]{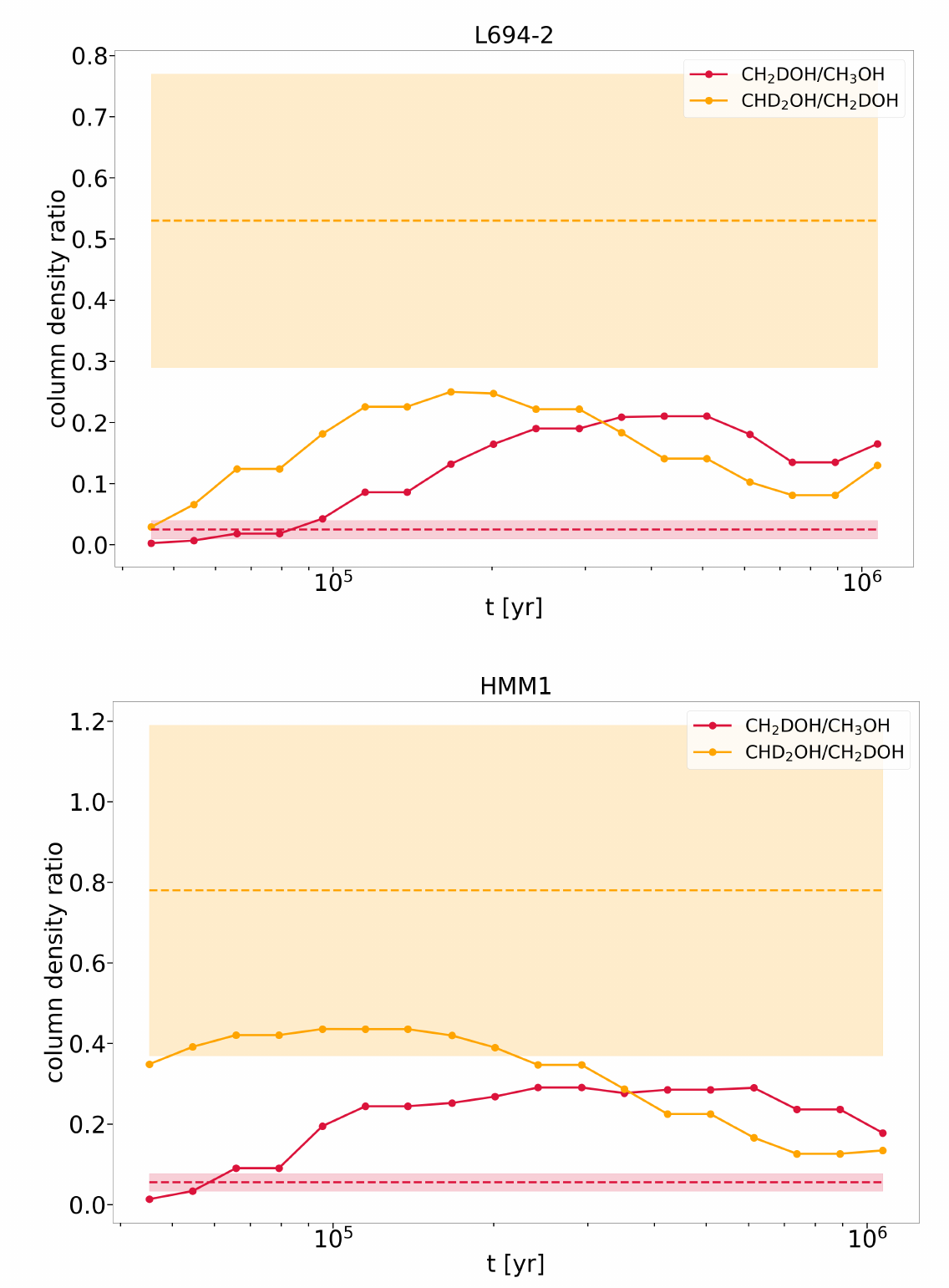}% This is a *.eps file
\label{fig:other_cores}
\end{center}
\caption{Results from the best model (D5) from Riedel et al. (2025)\cite{riedel25} for the pre-stellar cores L694-2 and HMM-1. The horizontal dashed lines show the result from
the observations and the shaded region indicates the error bars of the observed ratio \cite{lin23a}.}
\end{figure}

\begin{figure}[!ht]
\begin{center}
\label{fig:profiles}
\includegraphics[width=14cm]{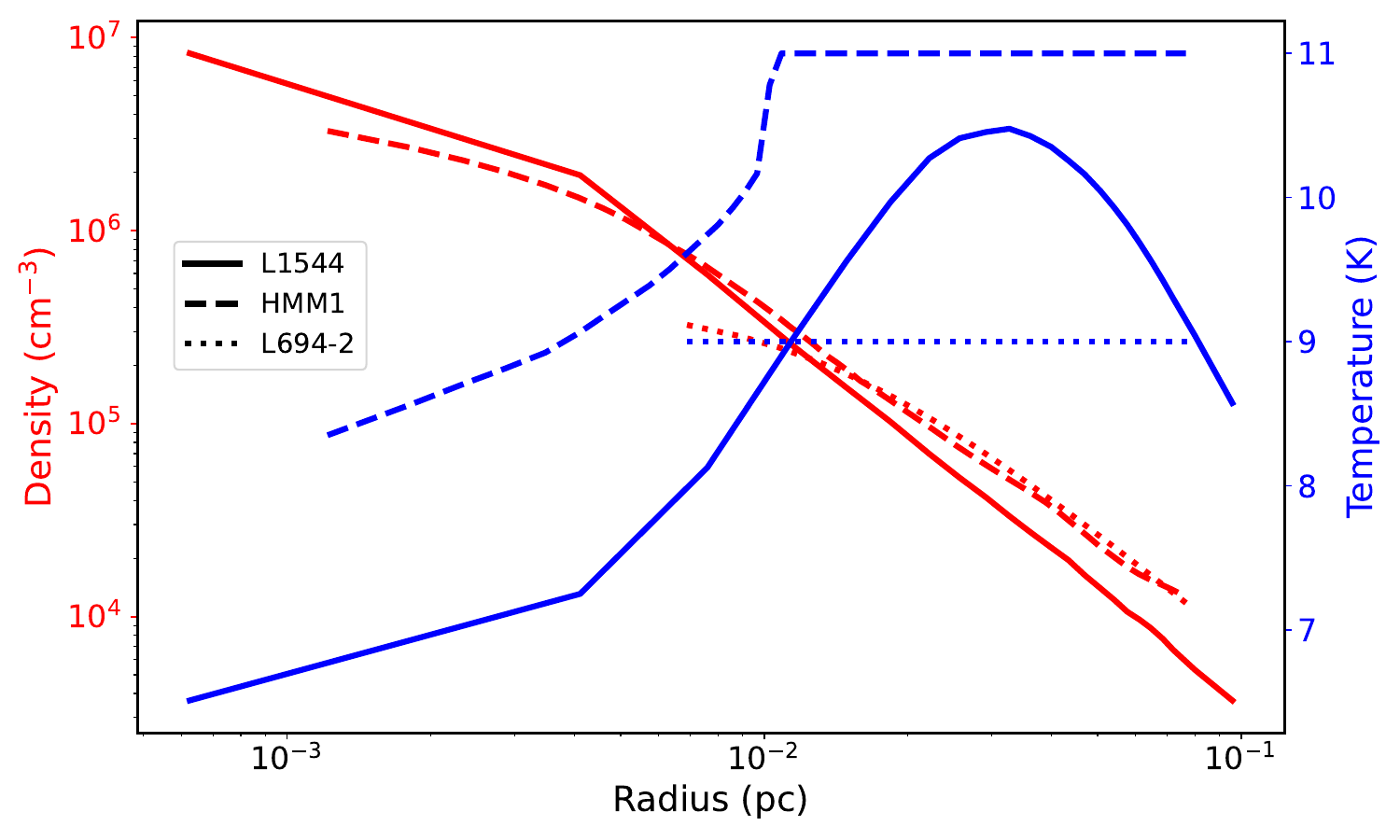}% This is a *.eps file
\end{center}
\caption{Physical structures of L1544 \cite{keto15}, HMM1 \cite{pineda22, harju24}, and L694-2 \cite{lin23b} used for the chemical models.}
\end{figure}

\begin{figure}[!ht]
\begin{center}
\includegraphics[width=14cm]{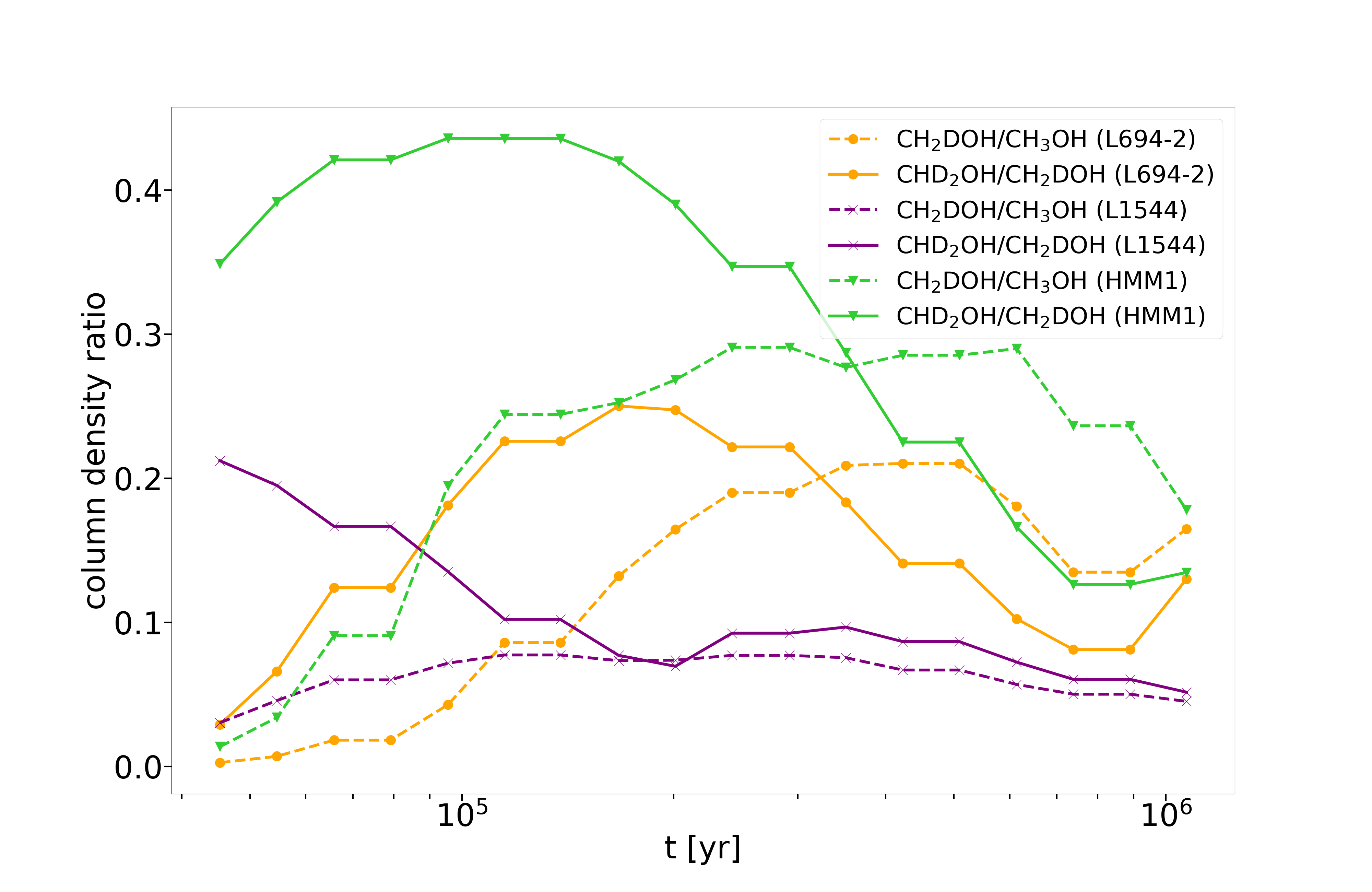}% This is a *.eps file
\end{center}
\caption{R$_{D}$ and R$_{D_2}$ for methanol predicted by the model D5 from Riedel et al. (2025) using the physical structures of L1544 \cite{keto15}, HMM1 \cite{pineda22, harju24}, and L694-2 \cite{lin23b}.}
\end{figure}

\clearpage
\section{Dependence of deuteration on physical conditions}

The dependence of the modelled R$_{\mathrm{D}}$ and R$_{\mathrm{D_2}}$ on the physical properties, mainly the H$_2$ density and temperature, is a complicated multidimensional problem and a detailed exploration is beyond the scope of this paper.

The fractionation of the deuterium in the gas phase is most efficient in the cold and dense center of the pre-stellar core. H and D atoms are directly adsorbed onto the surface of dust grains from the gas phase, where the atomic D/H ratio is enhanced. However, an increased atomic D/H ratio on the grain by itself does not guarantee, that the deuterium atoms are able to meet their reaction partners successfully. This also depends on the efficiency of the diffusion process, that is dictated by the employed mode of diffusion and the grain temperature. In general, higher grain temperatures allow for a faster diffusion process and a higher reaction rate of potential reaction partners.

\begin{figure*}[h!] 
    \centering
	\includegraphics*[width=0.8\textwidth]{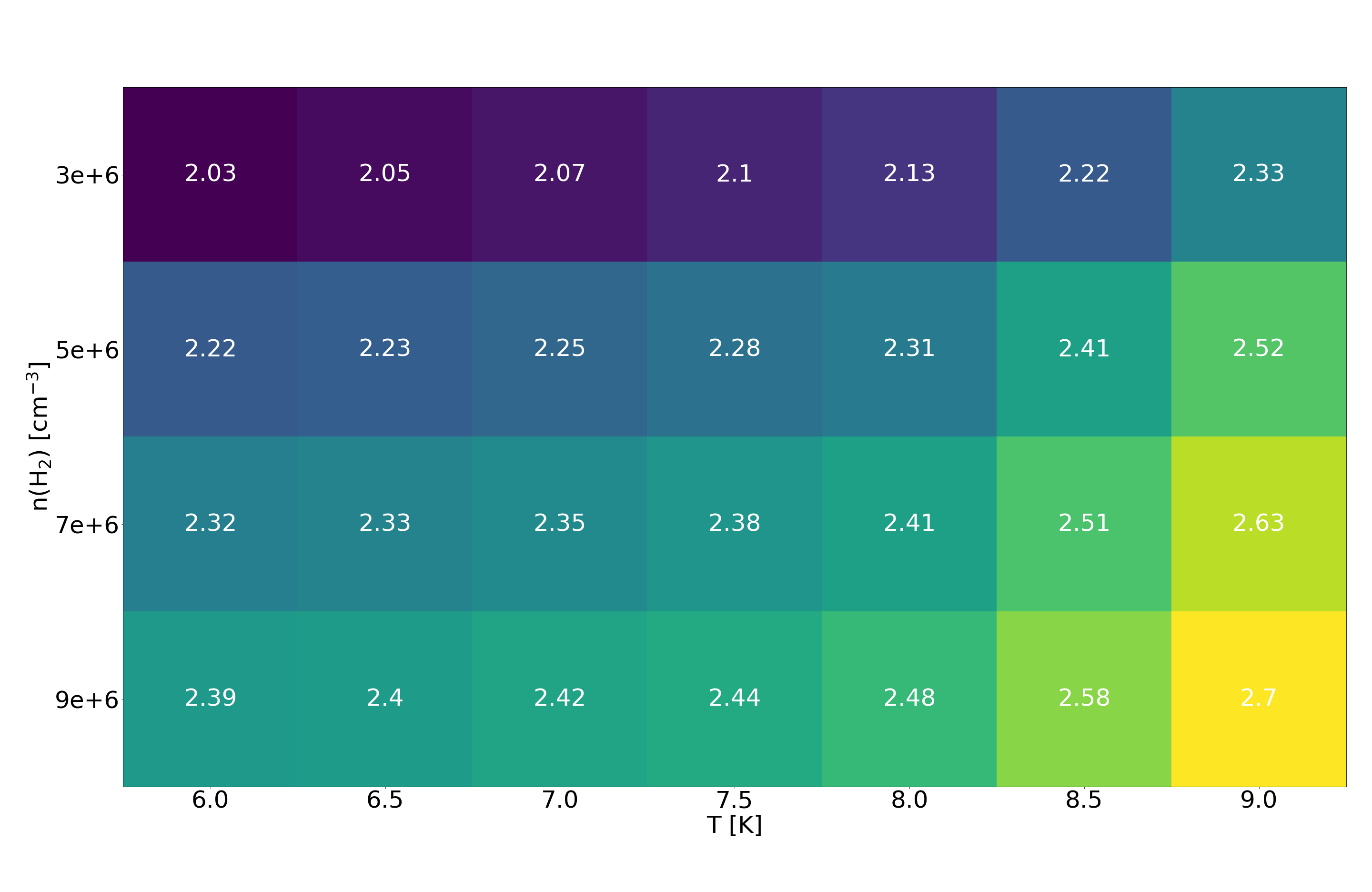}
    \caption{Heatmap of the abundance ratio of CH$_2$DOH over CH$_3$OH for a parameter grid with $n(\mathrm{H_2})$ between \SI{3e+6}{\per\cubic\centi\meter} to \SI{9e+6}{\per\cubic\centi\meter} and $T_{\mathrm{gas}}$ and $T_{\mathrm{grain}}$ between \SI{6.0}{K} and \SI{9.0}{K}.}
    \label{fig:heatmap_CH2DOH_CH3OH}
\end{figure*}

\begin{figure*}[h!] 
    \centering
	\includegraphics*[width=0.8\textwidth]{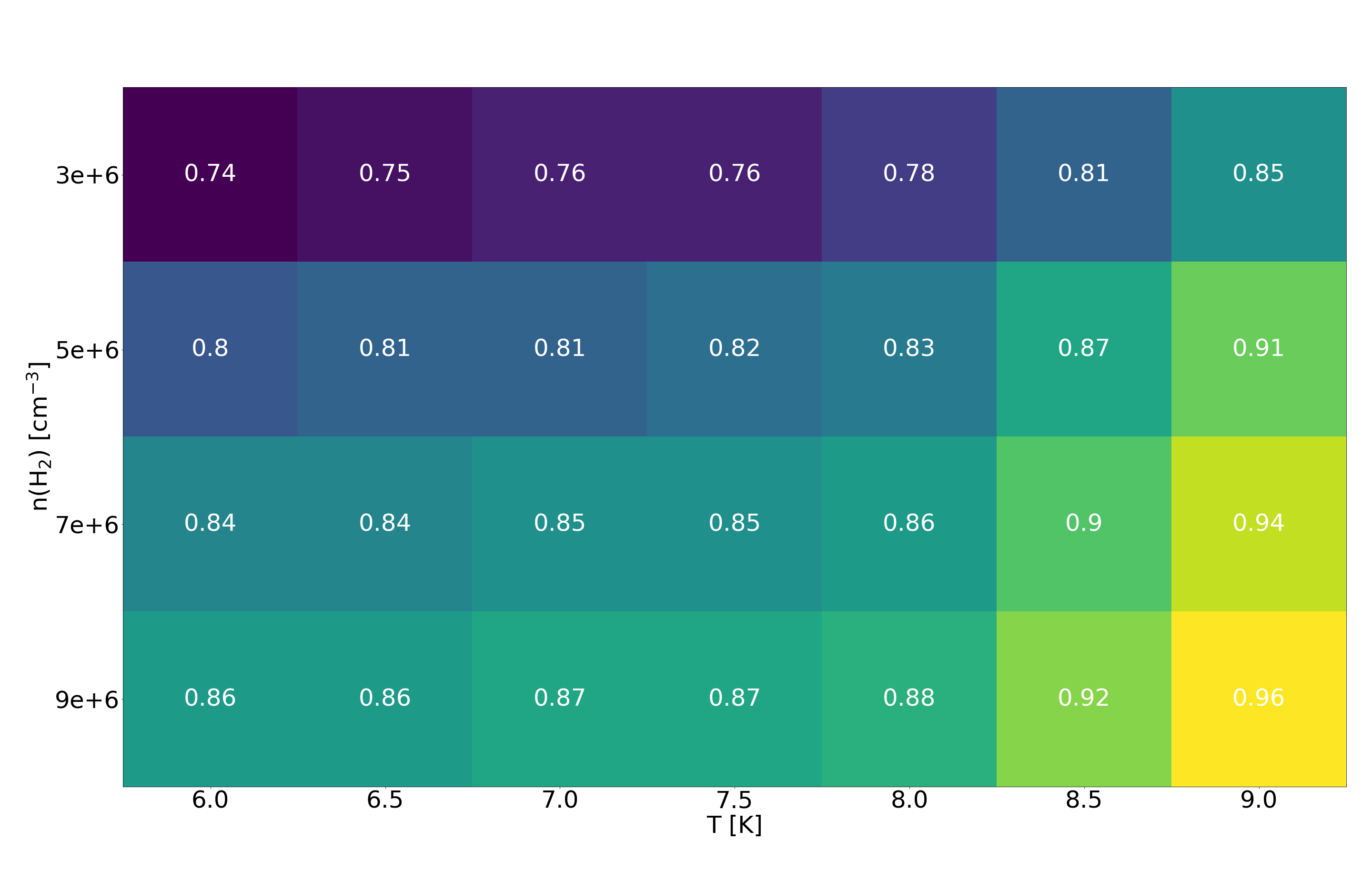}
    \caption{Heatmap of the abundance ratio of CHD$_2$OH over CH$_2$DOH for a parameter grid with $n(\mathrm{H_2})$ between \SI{3e+6}{\per\cubic\centi\meter} to \SI{9e+6}{\per\cubic\centi\meter} and $T_{\mathrm{gas}}$ and $T_{\mathrm{grain}}$ between \SI{6.0}{K} and \SI{9.0}{K}.}
    \label{fig:heatmap_CHD2OH_CH2DOH}
\end{figure*}

To shed some light into the density and temperature dependence, we have run a grid of static 0D simulations (see Figure \ref{fig:heatmap_CH2DOH_CH3OH} and \ref{fig:heatmap_CHD2OH_CH2DOH}). The H$_2$ densities range from \SI{3e+6}{\per\cubic\centi\meter} to \SI{9e+6}{\per\cubic\centi\meter}. The gas and grain temperatures are set to the same value and range from \SI{6.0}{K} to \SI{9.0}{K}. The selected range is an appropriate parameter range for the dust peak of the three pre-stellar cores (L1544, HMM1 and L694-2) investigated. This is where the highest levels of deuteration are expected to occur. All other parameters remain constant between individual runs. 

We conclude that both H$_2$ density and temperature affect the R$_{\mathrm{D}}$ and R$_{\mathrm{D_2}}$. The highest R$_{\mathrm{D}}$ and R$_{\mathrm{D_2}}$ ratios are determined for high densities $n(\mathrm{H_2})$ = \SI{9e+6}{\per\cubic\centi\meter}, where fast gas phase reactions promote an efficient fractionation process and short freeze-out timescales quickly deliver the atomic R$_{\mathrm{D}}$ to the grain's surface, and high $T_{\mathrm{grain}}$ = \SI{9}{K} speed up the diffusion process. We also note that partially similar R$_{\mathrm{D}}$ and R$_{\mathrm{D_2}}$ can be obtained by multiple parameter pairings.

\end{document}